\documentclass[prb,reprint]{revtex4-2}
    \setcitestyle{super}

\usepackage{lmodern}			
\usepackage[latin1]{inputenc}
\usepackage[T1]{fontenc}

\usepackage{graphicx}     % figures
\usepackage{amsmath}      % equations
\usepackage{tabularx}		  % tables
\usepackage{xcolor}			  % colored text

\usepackage{epstopdf}			% include eps-files
\usepackage{hyperref}   	% use for hypertext links
\usepackage[all]{hypcap} 	% Link jumps to the top of a graphic/table instead of jumping to its caption

\usepackage{blindtext}

\usepackage{todonotes}
\usepackage{tikz}

\usepackage{subcaption}	

\newcolumntype{C}[1]{>{\centering\arraybackslash}p{#1}}		% centered columns with fixed width, C{1cm}

\hypersetup{pdfstartview={XYZ null null 0.93}}
\hypersetup{pdfpagemode=UseNone}

\newcommand{\esub}{\epsilon_{\text{sub}}}

\newcommand{\zf}{\zeta_\text{f}}
\newcommand{\sif}{\sigma_\text{f}}
\newcommand{\drot}{D_\mathrm{rot}}
\newcommand{\etac}{\eta_\text{crit}}
\newcommand{\etat}{\eta_\text{trans}}

\begin{document}
\title{Interplay of orientational order and roughness in simulated thin film growth of anisotropically interacting particles}
\author{E. Empting}
\email{eelco.empting@uni-tuebingen.de}
\affiliation{Institut f\"ur Angewandte Physik, Universit\"at T\"ubingen, Auf der Morgenstelle 10, 72076 T\"ubingen, Germany} 

\author{N. Bader}
\affiliation{Institut f\"ur Angewandte Physik, Universit\"at T\"ubingen, Auf der Morgenstelle 10, 72076 T\"ubingen, Germany}

\author{M. Oettel}
\affiliation{Institut f\"ur Angewandte Physik, Universit\"at T\"ubingen, Auf der Morgenstelle 10, 72076 T\"ubingen, Germany} 

\begin{abstract}
	\noindent
 Roughness and orientational order in thin films of anisotropic particles are investigated using kinetic Monte Carlo simulations on a cubic lattice. Anisotropic next-neighbor interactions between the lattice particles were chosen to mimic the effects of shape anisotropy in the interactions of disc- or rod-like molecules with van-der-Waals attractions. Increasing anisotropy leads first to a preferred orientation in the film (which is close to the corresponding equilibrium transition) while the qualitative mode of roughness evolution (known from isotropic systems) does not change. At strong anisotropies, an effective step-edge (Ehrlich-Schwoebel) barrier appears and a non-equilibrium roughening effect is found, accompanied by re-ordering in the film which can be interpreted as the nucleation and growth of domains of lying-down discs or rods. The information on order and roughness is combined into a diagram of dynamic growth modes.       
 %Thin film growth of cubic particles is investigated using kinetic Monte Carlo simulations. Inter-particle interactions were anisotropic with respect to the relative particle positions, where the anisotropy of interactions was chosen to mimic the effects of shape anisotropy in a system of disc-shaped particles. The particle-substrate interaction was chosen to mimic growth of the film onto a substrate of different material. We find that the interaction anisotropy leads to an ordering transition, while due to the isotropic shape of the particles we can still observe different growth modes depending on the substrate material. At strong anisotropies, we find a non-equilibrium roughening effect, accompanied by re-ordering. We then tie these findings together in a non-equilibrium phase diagram for the roughness and ordering behavior.
\end{abstract}
\maketitle
\section{Introduction}
%\tb{More general first paragraphs: film growth, generic modelling to obtain general results on roughness evolution, height correlations...lattice SOS models have been quite useful. Citations. Roughness is a central variable: comparatively easy to measure and
%compute, desirable to control. We revisited this model, looking at the effect of substrates on the growth modes of thin films.} 

Film growth of organic molecules is a topic of considerable practical interest, since it is an indispensable tool in the fabrication of organic optoelectronic devices\cite{bruetting_2012}. The morphology of such films, notably manifested in the film roughness or the orientation of molecules in the film, significantly influences the electronic properties of the resulting device\cite{yang_yan_2009}. Besides the practical interest, film growth constitutes a basic non-equilibrium process for thermal many-body systems and is therefore of great theoretical relevance for the field of non-equilibrium statistical physics. Here, a large body of work has gone into the study of film growth with isotropic particles as realized in the growth of metal films. Roughness evolution is determined by the control parameters temperature, deposition speed and choice of substrate material \cite{evans_thiel_bartelt_2006}, and much insight and 
general results %on roughness evolution, height-height correlations etc. 
have been obtained using solid-on-solid (SOS) simulation models of particle growth on predefined lattices (very often taken to be a simple cubic lattice) 
\cite{vvedensky_1988,kaski_1991,binder_1997,raible_2002,reis_2020,reis_2021,martynec_2019,martynec_2021,pierrelouis_2007, pierrelouis_2010, pimpinelli_villain_1998, michely_krug_2004}. In recent work by our group, the transitions between different, basic modes of roughness evolution (layer-by-layer growth, island growth, rough growth from the start) which occur upon changing control parameters or happen as a function of the amount of deposited material, were studied\cite{empting_2021}.    

%It is especially desirable to control roughness evolution, since this is significantly influenced by growth conditions such as temperature and substrate material\cite{evans_thiel_bartelt_2006}. In order to be able to better understand how growth conditions influence structural properties, simple solid-on-solid (SOS) simulation models have been successfully applied to obtain general results on roughness evolution, height-height correlations etc.\cite{kaski_1991,binder_1997,raible_2002} In a previous publication\cite{empting_2021} we employed kinetic Monte Carlo (KMC) simulations in order to study the effects of substrate interactions on the evolution of the film roughness. We found that, depending on the inter-particle interaction strength $\epsilon$, the substrate attraction $\esub$, and the kinetic parameters, the film will show different short- and medium-term growth behavior. The particles might initially form islands before coalescing into a smooth or rugged film, the film can grow in a layer-by-layer fashion, or the film might grow in a roughening way from the start.

The workhorse SOS model on cubic lattices can be extended in several ways: One can investigate a system of multiple particle species \cite{reis_2020_binary}, different lattice types \cite{evans_thiel_bartelt_2006}, or introduce anisotropic particles. Anisotropy may be split into shape and interaction anisotropy.
Here we mean by shape anisotropy that the steric core of particles is different from a sphere (continuum) or different from a cube (for models on cubic lattices). In contrast, interaction anisotropy refers to particles having an isotropic excluded volume interaction, but residual (attractive) interactions depend on the orientation of the particles. An example would be e.g. spherical particles with electric or magnetic multipole moments. 
The study of particles with anisotropic steric shapes has a long history in investigations of the behavior of liquid crystals\cite{martinez_raton_2014, quiring_2019}, while interaction anisotropy has been used to both study simple models of liquid crystals\cite{lebwohl_1972,luckhurst_1980} and the assembly of magnetic nano-cubes\cite{glotzer_2007} and to mimic effects of anisotropic substrates on film growth\cite{martynec_2018}. 
Growth with organic molecules in general means that both shape and interaction anisotropy are involved. An exception is the growth of films with C$_{60}$ \cite{bommel_2014, janke_2021}.
%There have also been studies in which both anisotropies were implemented\cite{huang_2011} \tr{Clancy}. 

\begin{figure}[h]
	\centering
	\includegraphics[width=0.9\linewidth]{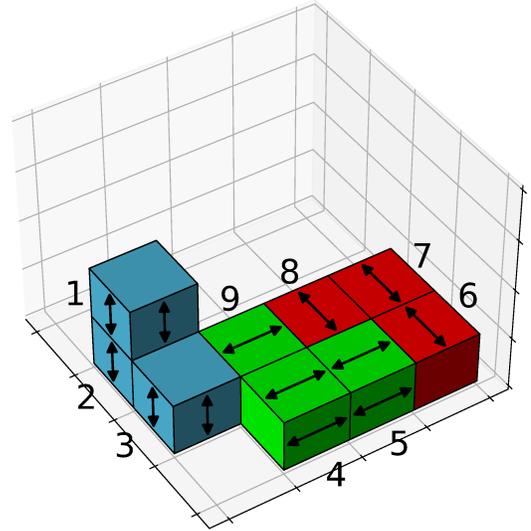}
	\caption{Schematic representation of nearest-neighbor interactions. Colors and double arrows indicate particle orientations. Particle pairs 1-2, 4-5, and 6-7 each interact with strength $-\eta \cdot \epsilon$, since the particles in each pair are aligned along their respective orientational axes. All other particle pairs interact with strength $-\epsilon$. Additionally, particle in the lowest layer interact with the substrate with a strength $-\esub$. Interactions are in units of the thermal energy $k_\text{B} T$.}
	\label{fig:interaction_schematic}
\end{figure}
In this work, we study by simulations the film growth with cubic particles possessing interaction anisotropy on cubic lattices. Particles are characterized by an orientation in either the $x$, $y$, or $z$ direction, and if two particles of the same orientation are aligned in this orientation direction, their interaction is stronger by a factor of $\eta$. This particular form of anisotropy was chosen to mimic the shape of disc-like organic molecules, e.g. benzene or perylene (see the discussion in Sec.~\ref{subsec:aniso} below). The fact that particles still only occupy one lattice site allows us to implement the solid-on-solid restriction, which prevents cavities inside the film and significantly reduces computational overhead. Despite the simplicity of the model, it depends on a number of parameters (particle-particle interaction, interaction anisotropy, substrate-particle interaction, lateral and rotational diffusion, growth rate) and we attempt a systematic study of the dependence on those parameters. The main focus is on the influence of the interaction anisotropy and orientational order in the film on the simple growth modes identified in Ref. \citenum{empting_2021} in the isotropic case. Previous work regarding lattice film growth has usually focused on either the sub-monolayer regime or on long-time scaling behavior. In contrast to this, we are interested in the evolution of roughness and orientation during the growth of a ``medium-thick'' film. We find that for small interaction anisotropies, there is a transition to orientational order in the film which does not influence the roughness growth modes. For large anisotropies, however, there is a new roughening effect which is accompanied by an increasing amount of particles oriented perpendicularly to the substrate.  

%We find that, depending on the strength of the anisotropy, particle orientations will be either isotropic or oriented in fibers along the substrate. We extract the critical anisotropy for this transition and compare it to the values obtained in equilibrium bulk simulations. We also show that, for reasonably small anisotropies, the growth modes observed in Ref. \citenum{empting_2021} still occur independently of the ordering behavior. However, for large $\eta$ we find a non-equilibrium ``roughening '' effect which is accompanied by an increasing amount of particles oriented perpendicularly to the substrate.

The paper is organized as follows: In Sec. \ref{sec:model} the simulation model is introduced in more detail. In Sec. \ref{sec:equi_sims} results of equilibrium bulk simulations are presented, followed by results of film growth simulations in Sec. \ref{sec:sims}. Sec. \ref{sec:discussion} gives a summary and discusses the relation of this work to experimental results.  

\section{Methods}
\label{sec:model}
\subsection{Model parameters and observables}
The model implemented is an extension of the SOS model presented in Ref. \citenum{empting_2021}, for a schematic picture see Fig.~\ref{fig:interaction_schematic}. Particles in the lowest layer interact with the substrate with a strength $-\esub = E_\text{particle-substrate}/(k_\text{B}T)$.  Additionally, each particle now has a preferred orientation in either the $x$, $y$ or $z$ direction. If the orientations of two next neighbor (NN)  particles are the same as the vector connecting the neighbors, they interact with a strength $-\eta \cdot \epsilon$ (where $\eta \ge 1$ for all of the results below) and $\epsilon = |E_\text{particle-particle}|/({k_\text{B}T})$. Here, $E_\text{particle-substrate}$ and $E_\text{particle-particle}$ are interaction energies between a particle and the substrate and between two particles, respectively.
If the orientations do not fulfill this condition, they interact with a strength $-\epsilon$. 
(In Fig.~\ref{fig:interaction_schematic}, e.g. particle pairs 1-2, 4-5, and 6-7 interact with strength $-\eta\epsilon$ while the other next neighbor pairs interact with $-\epsilon$.)
We note that this type of interaction differs from those employed in the well-known Lebwohl-Lasher model\cite{bates_2001} for the nematic phase in liquid crystals, since in our model alignment of identically oriented particles is only favored along the axis of orientation (in the Lebwohl-Lasher model, particle pairs 2-3, 4-9, and 7-8 in Fig. \ref{fig:interaction_schematic} would also interact with a strength of $-\eta \cdot \epsilon$, since there equal orientation is favored regardless of the NN position). 

The KMC film growth simulations are performed on a cubic lattice of lateral sizes $L=256$ with periodic boundary conditions in the $x$ and $y$ direction ($z=0$ defines the substrate plane). In  Figs. \ref{fig:surf_rot_growth_modes}, \ref{fig:all_rot_growth_modes}, \ref{fig:surf_rot_rough}, \ref{fig:all_rot_rough} we show results for $L=1024$ which confirm the absence of a noticeable finite-size dependence of roughness and order in the film. %\tr{For representative parameters, additional simulations were performed with $L=1024$ in order to check the finite-size dependence of the film morphology.}
New particles were inserted into the system at a rate $F$ at a random position on top of the film, respecting the SOS condition (no overhangs). The time $\Theta$ is measured via the amount of deposited material, given in units of a filled monolayer. Particles already in the system can either hop to a neighboring site at a rate $D$ (with the restriction that the height difference between the two sites is not more than one lattice site) or change their orientation vector (rotate) at a rate $\drot$.
Both $D$ and $\drot$ are given below in units of the deposition rate $F$ since only the ratios of these kinetic rates matter. $D$ is equivalent to the surface diffusion coefficient for
a free particle on the substrate plane or a filled layer. 
%For colloidal diffusion, one usually assumes $D \propto T \cdot \exp(-E_D/k_BT)$, while for metal-on-metal diffusion one assumes $D \propto \exp(-E_D/k_BT)$. 
$\drot$ is defined such that for particles with $n$ independent orientations, the concentration $c_i$ of a particle species $i$ relaxes towards the equilibrium value $1/n$ as $c_i(t) = 1/n + (c_i(0) - 1/n) \cdot \exp \left( -n \drot t\right)$ (the time $t$ is in units of $1/F$).
In the implementation of rotations we distinguish two cases: 
Either all particles, including those buried in the bulk of the film, can rotate (henceforth called \textbf{Model B}), or only those at the surface (\textbf{Model S}). Model B is intuitive for the lattice model as such, since rotations are not sterically blocked. Model S would apply for anisotropically shaped particles which cannot rotate after they have been buried.
Hops or rotations are accepted with a probability $p = \text{min}(1,\exp(-\Delta E))$, where $\Delta E$ is the change in internal energy this move would cause (in units of $k_\text{B}T$). We note that all simulations shown were performed without an Ehrlich-Schwoebel barrier.%, i.e. $E_\text{ES}=0$

A central variable in thin film growth is the roughness of the film, which is defined as
\begin{equation}
	\sigma = \sqrt{\frac{1}{N} \sum_{i=1}^{N} (h_i - \bar{h})^2}
\end{equation}
where $N$ is the number of lattice sites in the substrate plane, $h_i$ is the height of the film at site $i$, and $\bar{h}$ is the mean thickness of the film. Furthermore, the average orientation of the film is of interest, described by the order parameter
\begin{equation}
	\zeta = \frac{N_z - \tfrac{1}{2} (N_x + N_y)}{N_x + N_y + N_z}
\end{equation}
frequently used in the study of liquid crystals. Here, $N_\alpha$ is the number of particles oriented in direction $\alpha$. $\zeta$ varies between -0.5 (all particles are oriented parallel to the substrate plane) and 1 (all particles are oriented along the substrate normal). For an isotropic system it is 0.
This observable can be either calculated for the film as a whole or separately for each layer, in which case we denote it as $\zeta_l$.

For the study of the orientational behavior of a bulk film in equilibrium, the following three order parameters are useful:
\begin{equation}
	\zeta^{(i)} = \frac{N_i - \tfrac{1}{2} (N_j + N_k)}{N_i + N_j + N_k}
\end{equation}
where $i,j,k \in \{x,y,z\}$ and the indices are pairwise different. Note that $\zeta^{(z)} =\zeta$.

%this scenario, since this slightly ``washes out'' the transitions we are interested in.
\subsection{Relation between interaction and steric anisotropy}
\label{subsec:aniso}
With our choice of the interaction anisotropy, we incorporate to some extent an attractive interaction anisotropy for anisotropic molecules. One may assume a van der Waals attraction proportional to the contact area between two molecules. Consider a disc-shaped molecule which is characterized by an orientation vector perpendicular to the disc. The attractive interaction between two such molecules is stronger if the discs are oriented ``back-to-back''. In continuum modelling, this is realized e.g. by
the Gay-Berne potential for Benzene molecules \cite{golubkov_2006}. 
Particle interactions with $\eta>1$ favor the formation of ``fibers'' of particles with equal orientation (stacks of discs oriented ``back-to-back'' if one adopts the perspective from above). Note that such fibers also occur in models of magnetic nano-cubes \cite{glotzer_2007,szyndler_2012}, in contrast to the interactions there our model has non-polar interactions.

With a slight modification, such a lattice model may also represent rod-like particles.
Rods are characterized by an orientation vector along the rod axis, here van der Waals attraction is stronger when the rods are aligned ``side-by-side''. In the lattice model, this corresponds to stronger interactions $-\eta\epsilon$ of identically oriented, neighboring particles where the next-neighbor direction is not the orientation direction of the particles. In Fig. \ref{fig:interaction_schematic} the pairs 2-3, 4-9, and 7-8 would have these stronger attractions.

To further mimic steric anisotropies in the disc-like case, we can also define an enhanced substrate interaction of particles oriented perpendicularly to the substrate as $-\eta \,\esub$. This favors lying-down discs as found experimentally for metallic substrates. Such a case will be briefly discussed towards the end of Sec.~\ref{sec:sims}. 
In the rod-like case, this enhanced substrate interaction energy would apply to $x$- and $y$-oriented particles in the first layer. 

% In KMC latticesimulations of explicitly anisotropic particles it is likely that cavity-rich films are grown due to particles sterically blocking each other. (Therefore, previous work with anisotropic particles on lattices concentrated on the monolayer regime \cite{klopotek_2017,clancy_2006}.) It also leads to a proliferation of parameters for the translational and rotational moves since they in general depend on orientation and pivot. Of course, such problems are avoided in atomistic simulations of film growth \cite{muccioli_2011,muccioli_2018} but in atomistic simulations parameter scans are difficult as they would correspond to simulations of a variety of different molecules where additionally deposition speed and temperature should be varied.   

We see the model setup as a first attempt to represent shape anisotropies solely by interactions. With regard to film growth of anisotropically shaped particles, studies so far have mostly employed molecular dynamics simulations \cite{muccioli_2011,muccioli_2018,ikeda_2019}, which are inherently limited in the system sizes and time scales which can be observed due to their high computational cost. On the other hand, the implementation of a lattice kinetic Monte Carlo (KMC) simulation of these particles poses major challenges regarding the implementations and rates of different moves when particles become elongated \cite{klopotek_2017}. Simulations of particles with an extension over 2 or 3 lattice sites have been undertaken\cite{clancy_2006,clancy_2007}, concentrating on the (sub-)monolayer regime. Thick films of dipolar particles extending over 2 lattice sites have been studied in Ref.~\citenum{huang_2011}, here understood as growth in a solvent (``colloidal growth''). In these works, a systematic study of film properties on the growth speed and interaction parameters was not undertaken. Off-lattice KMC simulations have also been performed\cite{neumann_2013}, but due to the higher number of degrees of freedom, these are again severely limited in the number of particles which can be simulated. 
Hence the restriction to interaction anisotropy is reasonable. 
%we performed these simulations in order to capture the effects of anisotropic shape solely by interaction anisotropy. }

%Therefore it is interesting to compare to the experimental literature with regard to orientation effects and roughness behavior and discuss the relation to the present work. 

%On the one hand, this makes solid-on-solid simulations impossible, while on the other hand it also creates the problem of properly implementing all possible rotation and translation moves. These factors lead to such simulations being computationally expensive and mostly limited to the deposition of few layers at very high fluxes. 
%However, this approach of anisotropic interactions combined with isotropic particles might lead to insights into film growth processes at longer timescales.
\subsection{Effective Ehrlich-Schwoebel barrier}
In our simulations, there is no explicit Ehrlich-Schwoebel (ES) barrier\cite{ehrlich_1966,schwoebel_1969}, $E_\text{ES} = 0$. ES barriers can be incorporated asymmetrically (only hops ``downwards'' have a reduced probability by a factor $\exp(-E_\text{ES})$) or symmetrically (hops both upwards and downwards have a reduced probability)\cite{leal_2011}.
%In simulations incorporating a symmetric ES barrier, downhill hops are favored due to increasing the amount of nearest-neighbor particles\cite{leal_2011}.
\begin{figure}[h]
    \centering
    \includegraphics{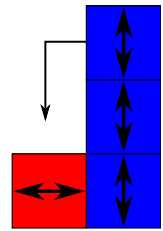}
    \caption{Schematic illustration for the effective Ehrlich-Schwoebel barrier. For the green particle it becomes energetically unfavorable to hop down when $\eta > 2$.}
    \label{fig:es_schematic}
\end{figure}
Interaction anisotropy leads to an effective, asymmetric ES barrier. Consider a $z$ particle (i.e. a particle oriented in $z$ direction) situated on top of another $z$ particle, as sketched in Fig. \ref{fig:es_schematic}.
The probability $p = \text{min}(1,\exp(-\Delta E))$ for
hopping down onto an $x$ or $y$ particle involves the energy change $\Delta E = 2 \epsilon - \eta \cdot \epsilon$ and is therefore smaller than 1 for $\eta > 2$, i.e. this hop becomes energetically unfavorable. In turn, the reverse move of hopping up is favored (its probability is 1), one can view this as the effect of an effective asymmetric ES barrier. As shown in Sec. \ref{sec:noneq} below, for large enough $\eta$ we find a strong roughening effect which can be traced back to this effective barrier. The exact value of $\eta$ at which this barrier becomes relevant depends strongly on the kinetic and energetic parameters, since it effectively depends on the average particle environment for downward hops.

\subsection{Remarks on parameter scaling}

We investigate $\epsilon$ in the range $2..6$, $D=10^3...10^5$, and $\drot=10...10^4$.
A direct comparison to experiments with organic molecules is complicated by the fact that the energetic and kinetic parameters are different from the parameters investigated here, in such experiments typically $\epsilon=10...15$ and $D=10^9...10^{11}$. 
However, there are indications that the use of smaller $\epsilon$ \textit{and} smaller $D$ leads
to similar morphological behavior. This holds e.g. for the island density in submonolayer growth in KMC simulations (here the original argument in Ref. \citenum{empting_2021} is incorrect, but
see a correction\cite{empting_corr} based on an analysis of the data of Ref. \citenum{martynec_2021}). In the absence of an ES barrier, a scaling relation for the roughness in 3D growth has been found in Ref. \citenum{aarao_reis_2015}. Both examples share a dependence on a scaling variable
$D^n(c+\exp(-\epsilon))$ where $n\ge 1$ and $c$ is some constant. Therefore, this scaling is
only useful for $\epsilon \lesssim - \ln c$ which limits the applicability in monolayer growth
to $\epsilon \lesssim 9$ in the monolayer case and $\epsilon \lesssim 4$ in the 3D growth case.
For more realistic 3D growth (in the isotropic case), the ES barrier plays an important role. 
Here, it has been demonstrated in Refs. \citenum{Trofimov2000} and \citenum{Trofimov2005} that
the transition between layer--by--layer and rough 3D growth depends on a scaling variable
$D^n \,\exp(-E_\text{ES})$. Unfortunately, scaling variables and relations involving all three variables $D$, $\epsilon$, $E_\text{ES}$ (for isotropic growth) are not known, and for anisotropic models like the present one the situation is even less clear. 
%Thus, these lower values of $\epsilon$ and $D$ were chosen to adequately model also the roughness behavior for experimental systems, if the scaling relations hold also for the anisotropic system. 
In particular, for our choice of $\drot$ there is no experimental justification; the choice of $\drot$ being smaller than $D$ corresponds to assuming (free) energy barriers for rotations to be higher than those for hopping moves.
%In Ref. \citenum{empting_2021} we argued that using low values of $D$ and $\esub$ and extrapolating to higher values works due to scaling of the morphology with these parameters (albeit the original argument was misguided; see corrections for Ref. \citenum{empting_2021}). As shown in Refs. \citenum{Trofimov2000} and \citenum{Trofimov2005}, the film morphology of multilayer films additionally scales with the Ehrlich-Schwoebel barrier, explaining the different growth modes at constant values of $\eta$ shown below.

 \subsection{Illustrative snapshots}
\begin{figure*}
\includegraphics[width=0.9\linewidth]{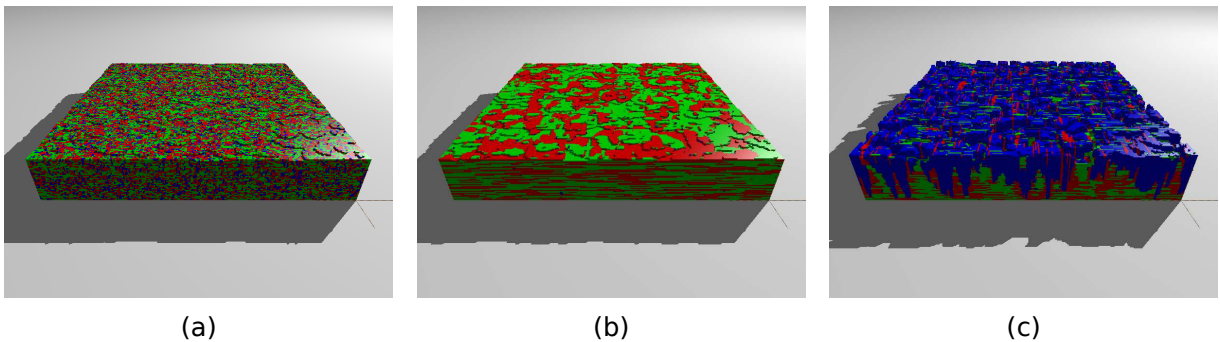}
\caption{Snapshots of films after deposition of 50 monolayers at $\epsilon=3$, $\esub=3.11$, $D=10^4$,$\drot=100$ in model S. The colors red, green, and blue indicate that particles are oriented in the $x$, $y$ and $z$ direction, respectively. From left to right $\eta$ increases from $1.5$ to $2.5$ and $3.5$. We see how the film first goes from disordered to ordered without the morphology changing. When increasing $\eta$ to 3.5, long fibers in the $z$ direction start forming, which lead to a significantly rougher film.}
\label{fig:snapshots}
\end{figure*}

In Fig.~\ref{fig:snapshots},  we show three illustrative snapshots from the KMC simulations to representatively highlight the main effects of anisotropy which will be discussed in more detail below.
Parameters differ only in the value of the interaction anisotropy parameter $\eta$. The other conditions are such that for the isotropic case $\eta=1$ nearly perfect layer-by-layer (LBL) growth is observed. 

In Fig.~\ref{fig:snapshots}(a), the anisotropy is low ($\eta$=1.5) and the film is isotropic and has grown LBL. With increased anisotropy ($\eta$=2.5 in Fig.~\ref{fig:snapshots}(b)), one observes a depletion of $z$--oriented particles (i.e. $\zeta \approx -0.5$) since particles may form energetically advantageous fibers of $x$-oriented and $y$-oriented particles ($x$-fibers and $y$-fibers, for short) in the film plane by diffusion. The domains composed of $x$- and
$y$-fibers (red and green in Fig.~\ref{fig:snapshots}) are small and themselves disordered, hence the fibers are relatively short. The film growth is still LBL. For even higher anisotropy  ($\eta$=3.5 in Fig.~\ref{fig:snapshots}(c)) longer $z$-fibers 
(blue in  Fig.~\ref{fig:snapshots}) have formed ($\zeta>0$) which cause a significantly rougher film and a clear deviation from LBL.     
The exemplary transitions in Fig.~\ref{fig:snapshots} (ordering and roughening) will be examined in more detail below.

\section{Equilibrium bulk phase behavior}
\label{sec:equi_sims}

For a thick bulk film, one expects an ordering transition with increasing $\eta$. Since the different orientations can be viewed as different particle species, the ordering transition corresponds to a demixing transition between the three species.  

In order to determine the location of the equilibrium ordering transition, we performed canonical Monte Carlo (MC) simulations in a completely filled box of size $L \times L \times L$ with $L=100$. The system was initialized with randomly oriented particles. A number $N_\text{step}$ of time steps is performed, during each of which a random particle was selected, a new random orientation was proposed, and then accepted with probability $p=\text{min}(1,\exp(-\Delta E))$, where $\Delta E$ is the change in internal energy this rotation would cause. This Metropolis step ensures that, eventually, the system will reach its equilibrium orientation.

 \begin{figure}[h]
    \centering
    \includegraphics[width=\linewidth]{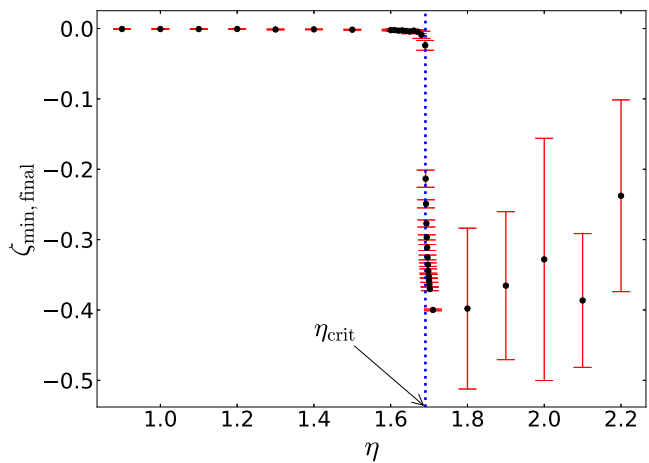}
    \caption{$\zeta_\text{min,final}$ vs $\eta$ in the canonical equilibrium system for $\epsilon=3$ . For $\eta < \etac$, the system is disordered and all $\zeta^{(i)}\approx 0$. For $\eta \ge \etac$, the system will start phase separating, leading to $\zeta_\text{min,final}<0$}
    \label{fig:q_min}
\end{figure}

For a given $\epsilon$, sweeps of $\eta$ are performed and for each value of $\eta$ the three order parameters $\zeta^{(x)}$, $\zeta^{(y)}$, and $\zeta^{(z)}$ are calculated after equilibration. We typically chose $10^{11}$ steps for equilibration and 10 independent runs for measurement. The value $\etac$ for the ordering transition is found by examining  $\zeta_\text{min,final} = \text{min}_i \zeta^{(i)}$ which exhibits a sharp jump at $\eta=\etac$.
This is shown in Fig. \ref{fig:q_min} for the choice $\epsilon=3$: here one observes the jump at $\etac \approx 1.69$ from a disordered system to one in which two particle species dominate and one is depleted ($\zeta_\text{min,final}<0$).
This is similar to the behavior observed in systems of hard rods on a lattice for intermediate rod lengths \cite{gschwind_2017,Vigneshwar_2017}. This effect of two dominating species is a consequence of the interactions favoring fibers. Consider planes with $z=\text{const.}$ in their state with lowest energy; these are completely filled with $x$-- or $y$--fibers (call them $x$-- and $y$--planes, respectively). There is no energetic difference between two neighboring $x$--planes or an $x$--plane next to a $y$--plane.  Entropically, however, it is favorable to allow different orientations in subsequent layers.
%The large variation in $\zeta_\text{min,final}$ after the transition can be explained by the fact that in this system, equilibration occurs extraordinarily slowly. Since we were not interested in the exact equilibrium values of $\zeta_\text{min,final}$, but only in the $\eta$ at which the transition occurred, we did not let the simulations run long enough for all systems to equilibrate.

\begin{figure}[h]
    \centering
    \includegraphics[width=\linewidth]{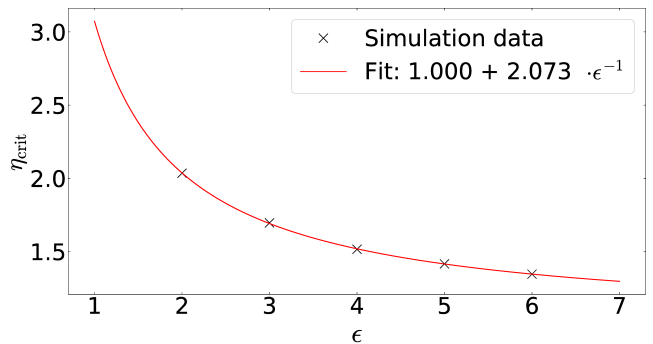}
    \caption{$\etac$ vs. $\epsilon$  in the canonical equilibrium system. The red full line shows a fit of the form $1 + \tfrac{\chi}{\epsilon}$.
    }
    \label{fig:etac_fit}
\end{figure}

From the MC simulations, the transition is found to be well described by 
\begin{equation}
   \label{eq:etac}
    \etac = 1 + \frac{\chi}{\epsilon}
\end{equation} 
with $\chi\approx 2.07$, as shown in Fig. \ref{fig:etac_fit}. In Sec. \ref{sec:sims} below, the $\etac$ found here will be compared to transition values found in growth simulations.

\section{{Non-equilibrium} Film growth simulations}
\label{sec:sims}

\subsection{Growth modes for isotropic interactions}
The first question of interest was whether the interaction anisotropy had any influence on the growth modes observed in Ref. \citenum{empting_2021}, obtained for isotropic particles. Depending on the parameters $D$, $\epsilon$, and $\esub$, four growth modes had been identified in Ref. \citenum{empting_2021}:  1) layer-by-layer (LBL), 2) forming islands at first, which then coalesce and form a smooth film (ISL$\to$LBL), 3) forming islands which coalesce into a non-smooth film (ISL$\to$CONST or ISL$\to$3D), and 4) growing a rough film from the beginning (3D). (Here, 3D growth refers to rough, three--dimensional growth and CONST refers to an approximate constant film roughness over time.) Note that for very long times (very thick films), all films roughen and thus show 3D growth \cite{aarao_reis_2015}. There is a rather sharp transition from growth mode 1 to mode 2 upon varying 
$\esub$ which is reminiscent of a dynamic wetting transition. Also, within growth mode 2 the change of island growth back to LBL growth happens at a rather sharp time (or coverage $\Theta$) and may be viewed as a ``flattening'' transition of the film.  

\subsection{Growth modes for anisotropic interactions}
\begin{figure}[h]
	\centering
    \includegraphics[width=\linewidth]{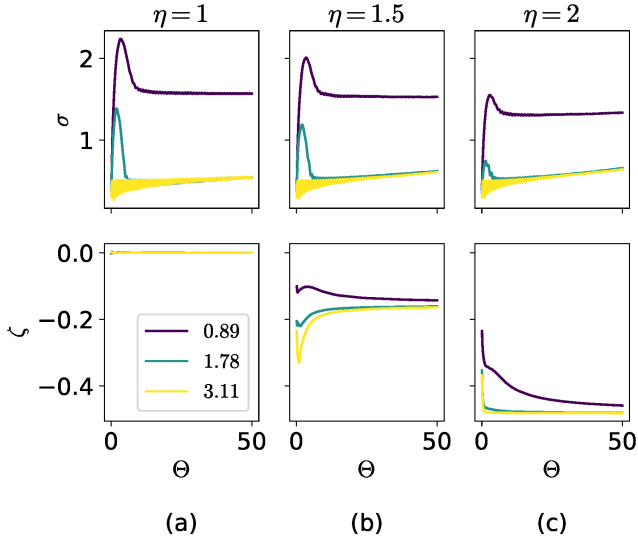}
	
	\caption{Roughness $\sigma$ and order parameter $\zeta$ vs $\Theta$ for $\epsilon=3$, $\Gamma = 10^4$, $\drot=100$ and $\eta = 1, 1.5, 2$ for different $\esub$ (different colors) in model S (for $L=1024$). The behavior of $\sigma$ vs $\Theta$ remains the same upon increasing $\eta$, while $\zeta$ changes from disordered to ordered. Note that $\eta=1$ corresponds to isotropy, so $\zeta=0$ is expected.}
	\label{fig:surf_rot_growth_modes}
\end{figure}
\begin{figure}[h]
	\centering
	\includegraphics[width=\linewidth]{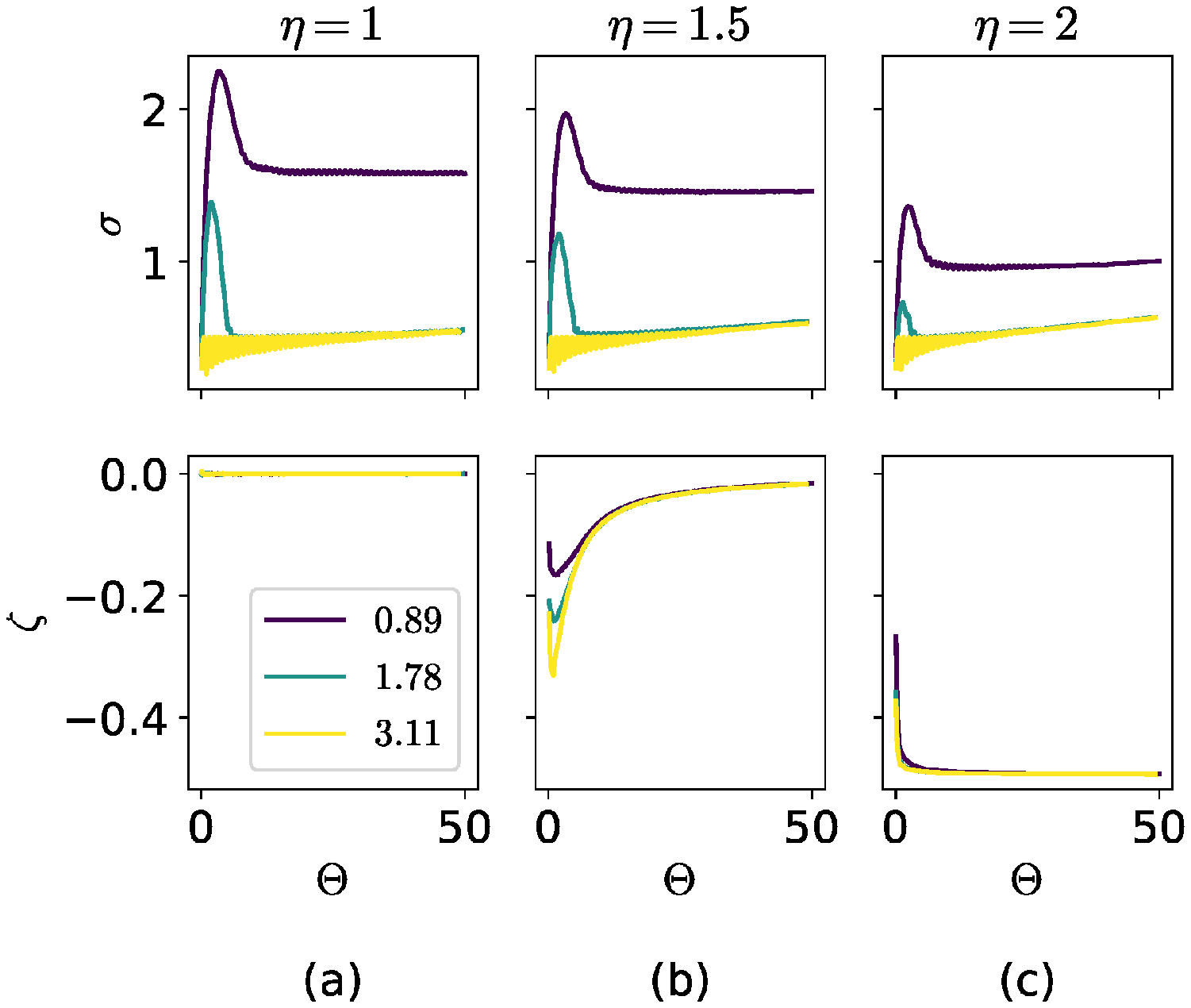}
	\caption{
	Same as Fig.~\ref{fig:surf_rot_growth_modes}, but for model B.}
	%Roughness $\sigma$ and order parameter $\zeta$ vs $\Theta$ for $\epsilon=3$, $\Gamma = 10^4$, $\drot=100$ and $\eta = 1, 1.5, 2$ for different $\esub$ in model B. The behavior of $\sigma$ vs $\Theta$ remains the same upon increasing $\eta$, while $\zeta$ changes from unordered to ordered.}
	\label{fig:all_rot_growth_modes}
\end{figure}
%\hline
%In Fig.~\ref{fig:snapshots} we show snapshots of typical films.In Fig.~\ref{fig:snapshots}(a), the anisotropy is low ($\eta$=1.5) and the film is isotropic and has grown LBL. With increased anisotropy ($\eta$=2.5 in Fig.~\ref{fig:snapshots}(b)), one observes a depletion of $z$--oriented particles (i.e. $\zeta \approx -0.5$) since particles may form energetically advantageous fibers of $x$-oriented and $y$-oriented particles ($x$-fibers and $y$-fibers, for short) in the film plane by diffusion. The domains composed of $x$- and
%$y$-fibers (red and green in Fig.~\ref{fig:snapshots}) are small and themselves disordered, hence the fibers are relatively short. The film growth is still LBL. For even higher anisotropy  ($\eta$=3.5 in Fig.~\ref{fig:snapshots}(c)) longer $z$-fibers 
%(blue in  Fig.~\ref{fig:snapshots}) have formed ($\zeta>0$) which cause a significantly rougher film and a clear deviation from LBL.

%In order to analyze these transitions, we show 
In Figs. \ref{fig:surf_rot_growth_modes} (model S) and \ref{fig:all_rot_growth_modes} (model B) we show typical examples for growth modes 1, 2, and 3 which hold for $\esub = 3.11,\ 1.78$, and $0.89$ respectively. The growth mode is read off the roughness curves (upper rows in Figs.~\ref{fig:surf_rot_growth_modes} and \ref{fig:all_rot_growth_modes}). LBL growth ($\esub=0.89$) shows an oscillating small roughness. ISL$\to$LBL growth ($\esub=1.78$) shows an initial sharp growth of roughness which collapses back to an oscillating small roughness.  ISL$\to$CONST growth ($\esub=3.11$) shows likewise an initial sharp growth of roughness which collapses back to a nearly constant roughness. Growth mode 4 is not shown here since it only occurs at very high fluxes or in the case of a non-vanishing Ehrlich-Schwoebel barrier.

The columns in Figs.~\ref{fig:surf_rot_growth_modes} and \ref{fig:all_rot_growth_modes} correspond to an increasing anisotropy parameter $\eta$ between 1 and 2 (from left to right). 
It is seen that upon increasing $\eta$, the system will not change its growth mode, even though the exact roughness values change. There is, however, a significant change in orientational order in the film (see lower rows in Figs.~\ref{fig:surf_rot_growth_modes} and \ref{fig:all_rot_growth_modes}, respectively). For the maximum value $\eta=2$, the final films in both model S and B are nearly perfectly ordered with $\zeta \approx - 0.5$ (all particles are oriented in either the $x$ or $y$ direction). 
For the intermediate value $\eta = 1.5$ it is seen that for model B the system will initially go towards an ordered state but  returns to disorder at higher coverages (Fig.~\ref{fig:all_rot_growth_modes}). For model S, a partial order remains (Fig.~\ref{fig:surf_rot_growth_modes}). 

This indicates that, for sufficiently weak anisotropies, the growth mode and the ordering of the film are independent of each other. Furthermore, for sufficiently thick films, the ordering is also independent of the strength of $\esub$ in both models, due to the short-ranged nature of the substrate attractions. The layer-dependent order parameter $\zeta_l$ is constant with increasing $l$ on all substrates for model B, while in model S it shows a gradient.

For weak substrates, $\esub=0.89$, in  Figs.~\ref{fig:surf_rot_growth_modes} and \ref{fig:all_rot_growth_modes}, the film grows in mode 3 (ISL$\to$CONST) and the constant roughness decreases upon increasing $\eta$. One would conjecture that in the limit of very high $\eta$, the growth mode of the system will then change to mode 2 (ISL$\to$LBL). We observe, however, that above a critical $\etat$, the film will no longer grow with all particles oriented in parallel to the substrate. Simultaneously, the roughness of the film will increase. This is a non-equilibrium transition (rapid roughening), as shown in Sec.~\ref{sec:noneq} below.

\subsection{Equilibrium transition}
\label{sec:eq}
\begin{figure}[h]
	\centering
	\includegraphics[width=\linewidth]{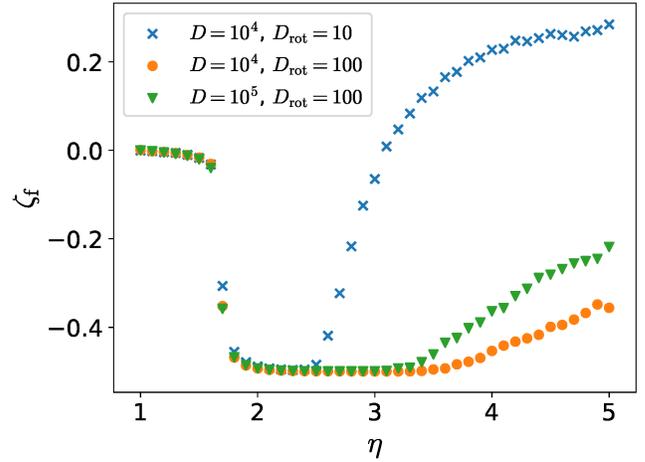}
	\caption{$\zf$ after deposition of 50 ML plotted vs $\eta$ for $\epsilon=3$ and $\esub=4$ at different $D$ and $\drot$ in model B. The first transition (from disordered to ordered) is independent of the chosen kinetic parameters.}
	\label{fig:equi_scaling}
\end{figure}
As shown in the previous section, upon increasing $\eta$ the system will initially go from an isotropic state to one in which all particles are oriented parallel to the substrate, independent of the strength of $\esub$. We suspect this to be an equilibrium transition similar to the one found in Sec. \ref{sec:equi_sims}, being independent of the kinetic parameters. 
%when plotting $\zf$ vs. $\eta$, where the location of $\etac$ is independent of the kinetics.
For model B, this is shown in Fig. \ref{fig:equi_scaling}, where the final order parameter $\zf$ (after deposition of 50 monolayers) vs $\eta$ is shown for different kinetic parameters. The first transition of $\zf$ vs. $\eta$ is indeed completely independent of the kinetic parameters in the investigated range $D=10^4 ...10^5$ and $\drot= 10..100$, indicating equilibrium behavior. 
After the transition the value of $\zf = -0.5$ indicates that all particles are oriented parallel to the substrate. Note, however, that this is only true in model B. In model S, the system can be frozen in an energetically less favorable state, leading to a $\zf$ which is larger than $-0.5$. 

\begin{figure}[h]
	\centering
	\includegraphics[width=\linewidth]{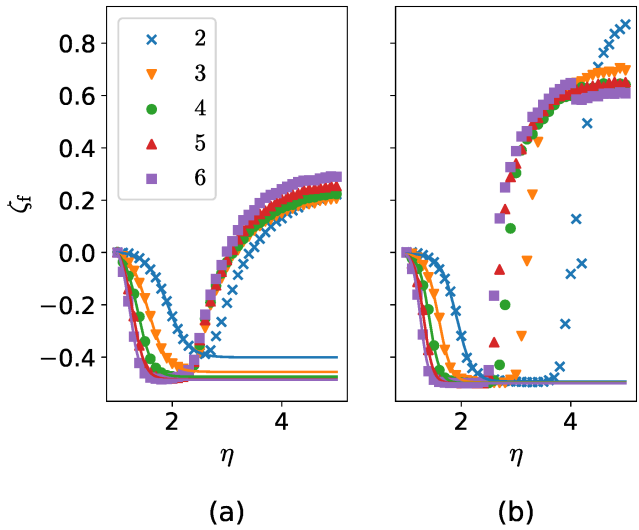}
	\caption{The order parameter $\zf$ after deposition of 50 ML vs $\eta$ at $\esub=4$, $D=10^4$ at (a) $\drot=10$ and (b) $\drot = 100$ for different $\epsilon$ in model S. The position of the first transition shifts with increasing $\epsilon$. The lines show the fitted tanh, the inflection point of which gives an estimate for the transition point.}
	\label{fig:surfrot_trans}
\end{figure}

\begin{figure}[h]
	\centering
	\includegraphics[width=\linewidth]{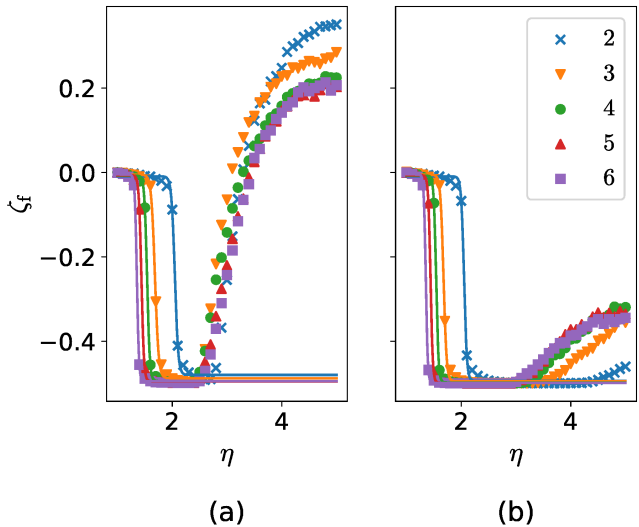}
	\caption{Same as Fig.~\ref{fig:surfrot_trans} but for model B.}
	%The order parameter $\zf$ after deposition of 50 ML vs $\eta$ at $\esub=-4$, $D=10^4$ at (a) $\drot=10$ and (b) $\drot = 100$ for different $\epsilon$ in model B. We can see how the position of the first transition shifts with increasing $\epsilon$. The lines show the fitted tanh, the inflection point of which we use as the point of the transition.}
	\label{fig:allrot_trans}
\end{figure}

The critical value $\eta$ for the disordered$\to$ordered transition is extracted from individual $\zf(\eta)$ curves by fitting
a tanh function to the data and extracting the inflection point as the transition point $\etac$ (see Fig. \ref{fig:surfrot_trans}  for model S and Fig. \ref{fig:allrot_trans} for model B).

\begin{figure}[h]
	\centering
	\includegraphics[width=\linewidth]{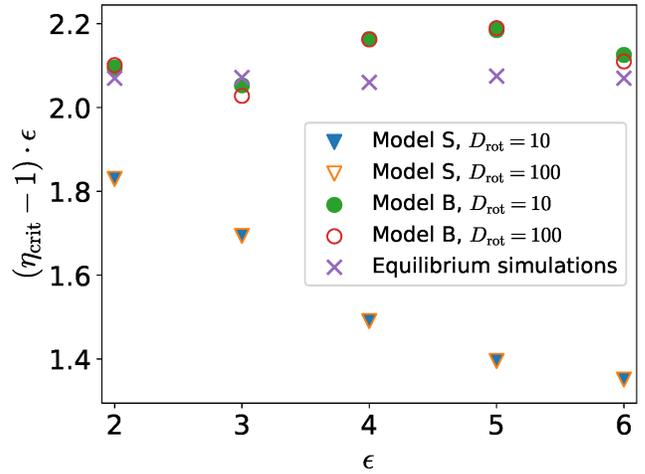}
	\caption{
		%Critical $\eta$ for the unordered to ordered transition for different $\epsilon$, compared to the mean-field result. We see that the transition point differs slightly between both models, which is expected due to the better equilibration in model B. The values we find for this system match those found in equilibrium canonical simulations very well.
		Critical $\eta$ for the disordered to ordered transition for different values of $\epsilon$ and re-scaled to $(\etac - 1) \cdot \epsilon$.
		%Values extracted from model B match those found in the equilibrium bulk simulations quite well. The values found in model B match those found in equilibrium simulations quite well, indicating that there the transition is an equilibrium one.
	}
	\label{fig:trans_pts_comp}
\end{figure}

In Fig. \ref{fig:trans_pts_comp} the extracted $\etac$ at different values $\epsilon$, re-scaled to $(\etac - 1)\cdot \epsilon$, is shown. In the equilibrium simulations, the re-scaled values are approximately constant, see Eq.~(\ref{eq:etac}). For model B, the data are close to this constant value, indicating that the transition is indeed the equilibrium one.
In model S, however, this scaling can no longer be found. Here $\etac$ is now significantly lower than the values found for model B. This can be explained by the fact that it is not necessary anymore to re-orient all particles in the film for the transition to occur, but only those at the surface. This leads to the transition occurring at lower anisotropies than in model B.

\subsection{Non-equilibrium transition}
\label{sec:noneq}

In Figs. \ref{fig:equi_scaling}, \ref{fig:surfrot_trans} and \ref{fig:allrot_trans}  it is seen that upon increasing $\eta$ beyond $\etac$, the final order parameter $\zf$ increases again. The critical $\etat$ for this transition does now depend on the kinetic parameters, indicating that this is a non-equilibrium transition. The increase of $\zeta$ is equivalent to the reintroduction of $z$--oriented particles into the system which for energetic reasons should be visible in the form of $z$--fibers in the system. Referring back to Fig.~\ref{fig:snapshots}(c), the formation of $z$--fibers is clearly visible. 

%As we have shown in the previous section, phase in which all particles are oriented in the $x$ and $y$ direction is only stable for smaller values $\eta$. If $\eta$ is larger than a critical $\etat$, we can see a change in both $\sigma$ and $\zeta$. The fact that the location of this $\etat$ depends on both $D/F$ and $\drot/F$ suggests that this is a non-equilibrium effect. Due to the nature of the implemented anisotropy, we expected these particles to preferably form ``fibers'' of equally oriented particles, given a large enough anisotropy.

\begin{figure}[h]
    \centering
    \includegraphics[width=\linewidth]{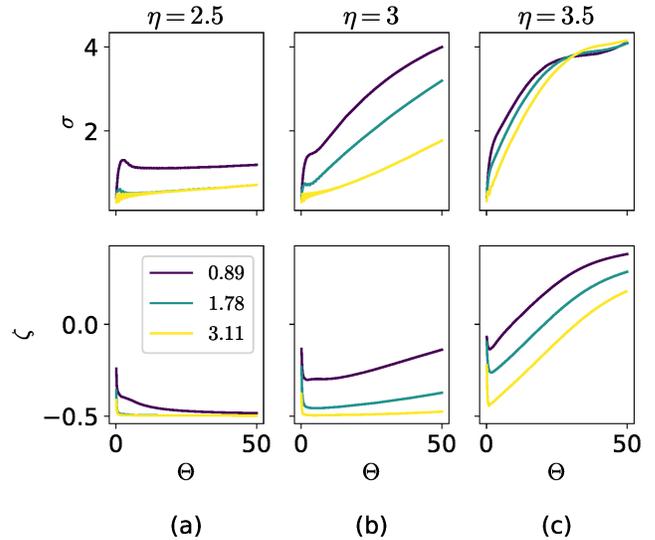}
    \caption{$\sigma$ and $\zeta$ vs $\Theta$ for $\epsilon=3$, $\Gamma = 10^4$, $\drot=100$ and $\eta = 2.5, 3, 3.5$ for different $\esub$ (different colors) in model S (for $L=1024$). Increasing $\eta$ past $\etat$ now leads to strong increases in both $\zeta$ and $\sigma$.}
    \label{fig:surf_rot_rough}
\end{figure}

\begin{figure}[h]
    \centering
    \includegraphics[width=\linewidth]{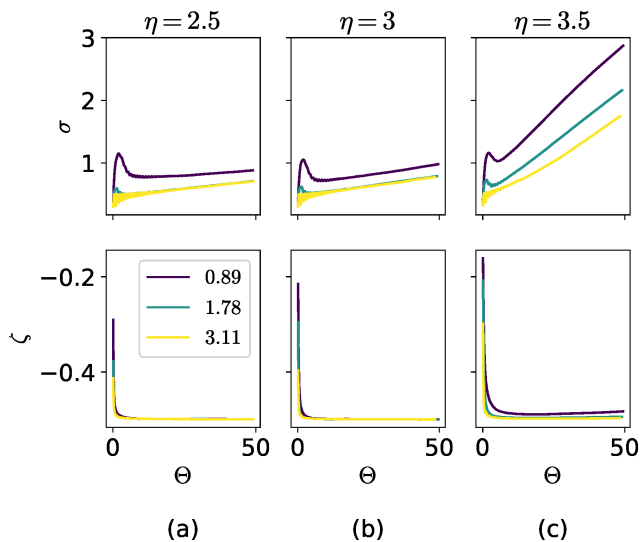}
    \caption{Same as Fig. \ref{fig:surf_rot_rough} but for model B}
    \label{fig:all_rot_rough}
\end{figure}

\begin{figure}[h]
	\centering
	\includegraphics[width=\linewidth]{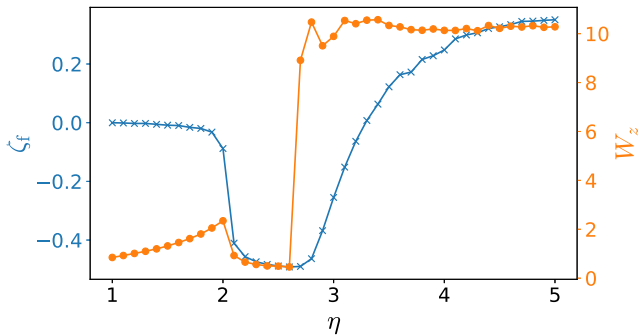}
	\caption{Comparison of $\zf$ and the width $W_z$ of the distribution of $z$ fibers vs $\eta$ for $\epsilon=3$, $\esub=4$, $D=10^4$, $\drot=100$ in model S. Both observables show transitions at approximately the same $\eta$ values. However, $\sigma$ shows a strong jump at the transition point, due to the onset of the formation of long fibers}
	\label{fig:trans_finder}
\end{figure}

In Figs. \ref{fig:surf_rot_rough} (model S) and \ref{fig:all_rot_rough} (model B) the behavior of $\sigma$ and $\zeta$ vs $\Theta$ is shown upon further increase of $\eta$ (these figures are a continuation of Figs.~\ref{fig:surf_rot_growth_modes} and \ref{fig:all_rot_growth_modes}). It is seen that the increase in $\zeta(\Theta)$ after the transition is accompanied with a substantial increase in roughness (rapid roughening). In both models the behavior of $\sigma(\Theta)$ and $\zeta(\Theta)$ strongly deviate from growth under weakly anisotropic conditions, suggesting that they might be used to pinpoint the $\etat$ for non-equilibrium growth. However, as shown in Fig. \ref{fig:trans_finder}, when plotting e.g. $\zf$ vs $\eta$, no clear transition point is visible, merely a gradual increase as $\eta$ is increased. The distributions of fiber lengths do, however, show a clear jump in behavior before and after the transition (see Appendix \ref{sec:appa}). In particular, the distribution of $z$--fibers shows a significant increase in the average length, in the maximum length and in the width of the distribution. After the transition, fibers can now grow through the whole thickness of the deposited film and reach a maximum length of $>50$ lattice sites.

Thus, we determined the exact transition point by running 5 simulations for each parameter set (in a system of $L=100$), measuring the lengths of all $z$--fibers, and choosing the $\eta$ at which the width $W_z$ of the $z$--fiber length distribution (excluding fibers of length $1$) jumps to a significantly higher value. This method proved robust enough to determine the critical $\eta$ for the non-equilibrium transition at a given $\epsilon$. Notable here was that the transition found through the jump in fiber length occurs slightly before the one we see in $\zf$ and that it is much sharper, indicating again that $\zf$ is not a good observable to detect such a non-equilibrium transition, since the relative amount of $z$--fibers is very small initially. Similarly, this also leads to the roughness $\sigma$ only changing slowly after the transition, making it equally unsuitable for determining the transition point.

\begin{figure}[h]
	\centering
	\includegraphics[width=\linewidth]{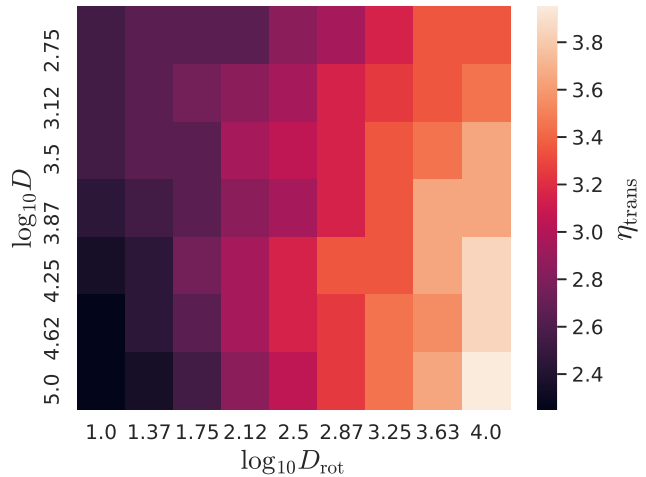}
	\caption{Heatmap showing the value of $\etat$ for the non-equilibrium roughening transition at different kinetic parameters and $\epsilon=3$ in model S. Even though the kinetic parameters are varied over several orders of magnitude, $\etat$ only slowly increases at higher $D$ and $\drot$}
	\label{fig:heatmap}
\end{figure}

We now swept the values of $D$ and $\drot$ over several orders of magnitude and pinpointed the $\etat$ for the non-equilibrium transition for both models B and S. In model B, the non-equilibrium transition could not be observed for $\eta$ up to $10^5$ when both $D$ and $\drot$ are large enough, indicating that such systems will always be in orientational equilibrium if particles can rotate fast enough. In model S, however, there will always be a non-equilibrium roughening transition. Fig. \ref{fig:heatmap} shows a heatmap of $\etat$ as function of $D$ and $\drot$ in model S. We see that, even though $D$ and $\drot$ are varied over several orders of magnitude, $\etat$ will only shift to slightly higher values. In general we can see that an increase in either $D$ or $\drot$ will increase the value of $\etat$. If $\drot$ is very small, however, the non-equilibrium transition is not very clear. This can be seen especially well for $\log_{10}\drot \lesssim 2$ in Fig. \ref{fig:heatmap}, where increasing $D$ does not necessarily lead to a higher $\etat$.

Putting together the findings of sections \ref{sec:eq} and \ref{sec:noneq}, one may construct a modified growth mode diagram in the $\esub-\eta$-plane for a given $\epsilon$. 

\begin{figure}[h]
	\centering
	\includegraphics[width=\linewidth]{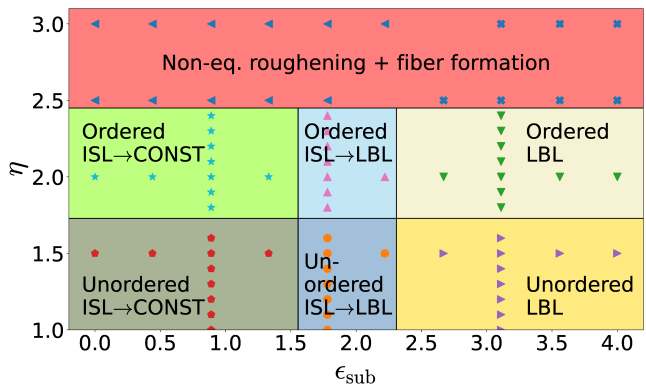}
	\caption{Modified growth mode diagram for $\epsilon=3$,$D=10^4$,$\drot=10$. Simulated system parameters are shown by symbols, colored regions display approximately the extent of the respective order--roughness growth mode.  For $\eta < \etat$, roughness mode and film ordering are independent of each other, leading to 6 kinds of film morphologies in this region. For higher values of $\eta$, the film will grow in a non-equilibrium manner, showing sudden changes in both roughness and order parameter}
	\label{fig:2d_phasediag}
\end{figure}

In Fig. \ref{fig:2d_phasediag}, this modified growth mode diagram is shown for $\epsilon=3$. The state of orientational order must be added to the roughness growth modes to describe a combined order--roughness growth mode. For $\eta \lesssim 2.4$ (below the non-equilibrium roughening transition) the roughness modes and film ordering are completely independent of each other. For stronger anisotropies, the system shows a new roughness growth mode (rapid non-equilibrium roughening) which is accompanied by $z$--fiber formation. 

Simulations with the parameters $\epsilon=4$, $D=10^4$, $\drot=10$
for three substrates $\esub=0.89, 1.78, 3.11$ confirmed the order and roughness transitions in this phase diagram. The transition values $\etat$ and $\etac$ are shifted to lower values. To cursorily check a possible dependence on $D$,  we performed simulations for $\epsilon=5$, $D=10^5$, $\drot=100$ with $\esub=4$ and varying anisotropy $\eta$. As in Fig.~\ref{fig:2d_phasediag}, the transition \textit{Unordered LBL $\to$ Ordered LBL $\to$ Non-equilibrium roughening}
has been found upon increasing $\eta$.

\subsection{Strong substrate interactions}

\begin{figure}[h]
	\centering
	\includegraphics[width=\linewidth]{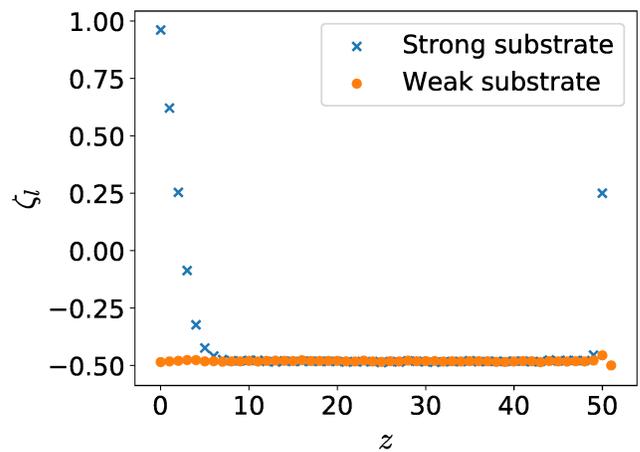}
	\caption{Layer--dependent order parameter $\zeta_l$ vs layer $z$ after deposition of 50 ML in model S for systems in which substrate interaction is (1) $\esub$ for all orientations (weakly interacting substrate) and (2) $\eta \,\esub$ for $z$-oriented particles (strongly interacting substrate), at $\epsilon=3$ and $\eta=2$. We see that on the strongly interacting substrate, the particles are all $z$-oriented at the substrate, and $\zeta_l$ slowly relaxes towards a state of only $x$- and $y$-oriented particles in higher layers. On the weakly interacting substrate, there is no change in ordering from lower to higher layers.}
	\label{fig:zeta_vs_height}
\end{figure}
\begin{figure}[h]
	\centering
	\includegraphics[width=\linewidth]{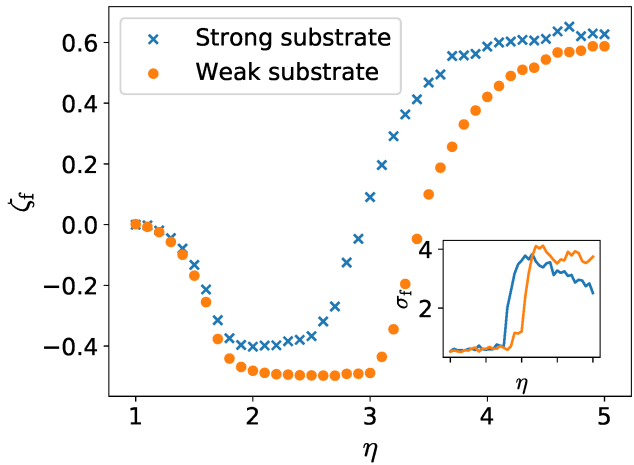}
	\caption{Comparison of total order parameter $\zeta$ in the film after deposition of 50 ML vs. $\eta$ between strongly and weakly interacting substrates for model S. 
	%for systems in which substrate interaction is (1) $\esub$ for all orientations and (2) $\eta \cdot \esub$ for $Z$ oriented particles. 
	The disordered $\to$ ordered transition occurs at the same position in both systems, but non-equilibrium roughening on the strongly interacting substrate is shifted to lower $\eta$ due to the first adsorbed layer consisting only of $z$-oriented particles. This earlier transition can also be seen in the behavior of $\sif$ (see inset). After the transition has occurred in both systems, however, $\sif$ is lower in the system grown on the strongly interacting substrate.}
	\label{fig:strong_weak_comp}
\end{figure}

So far, all simulations were performed on a substrate which interacts equally strongly with particles of all orientations (``weakly interacting'' substrate). We may define ``strongly interacting'' substrates by setting the substrate interaction of $z$-oriented particles to $-\eta \, \esub$, which reflects substrates preferring e.g. disk--like particles lying down flat. (In actual growth experiments, metallic substrates are of this kind and for those, surface orientation effects appear to be important; see Sec. \ref{sec:discussion} and Ref. \citenum{duerr_2003}) As the example for a moderate anisotropy of $\eta=2$ in  Fig. \ref{fig:zeta_vs_height} shows,  particles in the lowest layer will all be oriented in the $z$-direction, slowly relaxing to the equilibrium orientations seen in previous systems in higher layers. As a consequence, the total order parameter in the film $\zeta_f$ does not reach the equilibrium value of $-0.5$ (see Fig. \ref{fig:strong_weak_comp} ).

%Due to this behavior, we determined the transitions only using the weaker substrates. 

%However, the behavior found in this model is reminiscent of growth on substrates where one does expect the particle-substrate interaction to be particularly strong (see Discussion). For some cases, it might be necessary to implement this kind of anisotropy in order to accurately model an experimental system.
%However, in order to generate films which are more similar to real films (see Discussion), one should extend the interaction anisotropy also to $\esub$.\tr{TODO REALLY? Maybe more like: The argument can also be made that this is in fact the more realistic implementation, since  the scenario with the strong substrate will always lead to all particles in first layer orienting perpendicularly to substrate $\to$ more particles in layer than would be possible with steric repulsion. More extreme in the case of rod-like interactions. There we could only get a maximum of $L \cdot L / \eta$ particles lying in a layer, each interacting with $\eta \cdot \esub$. With strong sub, the max. sub energy can now be $L \cdot L \cdot \eta \cdot \eta$}

Additionally, it is seen that the non-equilibrium roughening transition is shifted to lower values of $\eta$ upon switching from the weakly to the strongly interacting substrate. This can be explained by the fact that the lowest layers are filled with $z$-oriented particles, acting as nucleation site for growing $z$-fibers which cause the roughening. Consequently, the new roughening growth mode found for anisotropic particles is further enhanced by strongly interacting substrates.

\subsection{Rod-like interactions}
\label{sec:rods}
We also investigated systems where the interactions were similar to those of rod-like molecules, i.e. NN particles interact strongly if their orientation is identical, but their positions are not aligned in the direction of the orientation. In Fig. \ref{fig:interaction_schematic}, particle pairs 2-3, 4-9, and 7-8 interact with strength  $-\eta\, \epsilon$, while all other particle pairs interact with a strength $-\epsilon$.

In these systems the same behavior was found as for disc-like interactions, only ``inverted''. This means that, upon increasing $\eta$ from $1$, the system will first go from an disordered state to one in which all particles are oriented in $z$- direction (i.e. $\zeta=1$) (standing-up rods). Upon further increase of $\eta$, again a non-equilibrium roughening transition is found, which is accompanied by more particles re-orienting in $x$ and $y$ directions (lying-down rods). For exemplary snapshots and further results for roughness and orientational order in the film,  see Appendix \ref{sec:appb}.

%We do, however, suspect that these kinds of molecules cannot be accurately modeled by this system.
We point out that the $z$-orientation in the film (standing-up rods) here is caused by energetic anisotropy, while for real rod-like molecules one expects strong entropic effects as well.
In simulations in which rod-shaped particles (attractive or not) were investigated, it was found that steric repulsions play a crucial role in the ordering behavior, leading e.g. to particles lying down on the substrate and standing up upon further filling of the first layer
(entropic standing--up transition \cite{klopotek_2017}).

\section{Summary and discussion}
\label{sec:discussion}
We have investigated a lattice SOS model for film growth with anisotropic particles where the anisotropy was modeled via next neighbor attractive interactions which depend upon the orientation of the particles (interaction anisotropy). 

The particular choice of interactions was made to (partially) capture the effect of interaction anisotropy for van der Waals-like interacting particles with an anisotropic shape. A new kinetic parameter in such a system is the rotation rate of particles $\drot$. We distinguished the two cases of model B (all particles in the growing film can rotate) and model S (only surface particles can rotate). If the anisotropy (with associated parameter $\eta$) is small,     
%We have seen that in a film-growth system where we emulate anisotropic particles via anisotropic interactions, 
the roughness growth modes identified in an SOS model for isotropic particles \cite{empting_2021} are preserved.  
In this regime we find upon increasing the parameter $\eta$ (for $\eta < \etat$) that the roughness evolution of the film remains the same as for $\eta=1$, and the particles in the film become oriented parallel to the substrate. For model B, this orientation transitions is independent of the kinetics of the growing film and takes place at the same value of $\eta$ as in an equilibrium system. This is somewhat different in model S, but also here the orientation transition is weakly depending on the kinetics and close to the equilibrium one. This explains the decoupling of roughness and orientation evolution for both rotation models. 
%We have seen that in a film-growth system where we emulate anisotropic particles via anisotropic interactions, the roughness growth modes which are observed in a ``normal'' SOS system are preserved when the anisotropy is small. 
%Upon increasing $\eta$, we find that initially, while the roughness evolution of the film remains the same, the particles in the film will now orient parallel to the substrate in what appears to be an equilibrium transition. When only particles at the surface can rotate, this transition is very smeared out and shifted to lower values of $\eta$. Comparison to canonical simulations show that in model B, however, this transition occurs at almost the same $\eta$, indicating that there the transition is indeed an equilibrium one.

For strong anisotropies ($\eta > \etat$), the growth mode is modified. The film will show rapid, non-equilibrium roughening due to the more pronounced occurrence of particles oriented perpendicularly to the substrate forming long fibers, leading to a rougher film and an increased order parameter.
In model B, this transition disappears at sufficiently high $D$ and $\drot$ (corresponding to slow growth), since in those cases particles can equilibrate fast enough to reach the (meta-)stable state of parallel orientation adopted in the initial growth. In contrast, in model S, the film will always show this non-equilibrium transition, since buried particles cannot rotate and thus cannot reach their equilibrium orientation.

This strong roughening effect can be rationalized by an inhibition of downhill diffusion for $z$ particles located on top of other $z$ particles, given that the interaction anisotropy is strong enough (higher than $\eta > 2$). In  the extreme case with no restrictions on the maximum vertical hop distance, the inhibition of downward hops leads to extremely tall, thin needles of $z$ particles.

The orientation effects we found correspond to the interaction asymmetry of disc-shaped molecules. Actual examples used in organic thin film experimental studies are
%The  disc-shaped organic molecules might interact with each other during thin-film growth experiments, for example 
phthalocyanines (e.g. CuPC and ZnPC), benzene, and perylene. These particles will, given the right conditions, form needles or stacks, similar to particles in our simulations at strong enough anisotropies.
The growth direction and ordering of such needles also strongly depends on the substrate. Notable here are
amorphous substrates, such as SiO$_2$ (corresponding to a weakly interacting substrate in our work) and
ordered substrates, such as metal surfaces (corresponding to a strongly interacting substrate).

Experiments with phthalocyanines on an amorphous substrate have shown that the disc-like molecules stand upright on the substrate and form needles consisting of herring-bone like arranged particles growing in all directions parallel to the substrate\cite{bao_1997}. On metallic substrates, particles in the first layer will lie flat on the substrate, initiating needle-growth perpendicular to the substrate, see \cite{haenel_2004} for perylene growth on copper or\cite{beyer_2014,ruffieux_2002} for HBC on Cu(111) and Au(111). These results are reminiscent of behaviors we found in simulations using weakly and strongly interacting substrates, respectively. For benzene and benzene derivatives it has been observed that, depending on the substrate material, benzene thin films will either form needles which are stable throughout the whole growth process\cite{yuan_2013} or they will form ordered structures in the first layer, but random crystallites in higher layers\cite{yuan_2013} (similarly for the derivative 1,3-bis(N-carbazolyl)benzene
(mCP) \cite{kwon_2013}). Such behavior we find here for particles with a relatively low anisotropy ($1 < \eta < \etac$) growing on strongly and weakly interacting substrates, respectively.
In Ref. \citenum{ishii_1996} it was found that when increasing the substrate temperature when depositing benzene onto gold-coated copper, the film ordering will go from amorphous (glassy) to partially crystallized to completely crystallized. This does not match well our findings that, for a given $\eta$, decreasing $\epsilon$ (increasing $T$) will move the film from a non-equilibrium roughened phase to an equilibrium phase (there is no glassy phase in our model). However, here the correspondence is not as clear since changing $T$ will also change $D$ and $\drot$.

%Due to the simple cubic arrangement of lattice sites we were unable to observe more complicated structures found in experiments\cite{Desiraju_1989_ActaCrystallogrSectBStructSci}, which limits the applicability of this model in predicting the exact structures which might occur during film growth. We can, however, show how growth conditions influence the general kinds of ordering and roughness of such a growing film.

An open question is whether needle-formation as described also leads to the kind of strong roughening we found in our simulations. Quantitative determinations of roughness evolution with time for disc-like molecules a scarce. An example for CuPC growth on SiO$_2$ seems to agree with this trend \cite{reisz_2021}. However, for a confirmation of orientation-dependent strong roughening one would need information on both roughness and the orientation in the film, otherwise strong roughening can be easily due to strong Ehrlich-Schwoebel barriers. A second example is the study in Ref.~\citenum{yang_2015} for F$_{16}$CuPc on  SiO$_2$ which shows a strong roughening after approx. 40 nm of film thickness ($\sim$ 27 layers of standing-up discs) which is clearly accompanied by a change in the type of surface grains but the orientation of the molecules after the strong roughening is unclear.

A wealth of data and observations can be found for thin film growth with rod-like molecules, most notably pentacene (PEN) and diindenoperylene (DIP). Here a strong roughening effect is found which is not seen in our model. For PEN and DIP growth on a weakly interacting substrate (SiO$_2$ at high enough temperature) there is initial LBL growth of standing-up rods for a few layers which changes into fairly strong 3D growth \cite{duerr_2006,zhang_2007,kowarik_2007,kowarik_2009,zhu_2011} and it does not seem to be accompanied by an appearance of lying-down rods. On the other hand, the appearance of islands of lying-down rods and ensuing roughening (as in our model, see App. \ref{sec:appb}) is seen for DIP on SiO$_2$ at lower temperatures \cite{duerr_2006} and for DIP on gold \cite{duerr_2003}. PEN does not show such an effect for SiO$_2$ at lower temperatures but a 1:1 mixture of PEN:PFP (perfluoropentacene) does \cite{hinderhofer_2011}. PEN on gold exhibits a lying-down monolayer and subsequent layers with predominantly standing-up molecules \cite{kaefer_2007}, similar to the order behavior shown in Fig.~\ref{fig:strong_weak_comp} but roughness data are not available.      
We have mainly discussed the relation to experimental results for growth of molecules in a vacuum chamber. The experimental conditions for growth in a colloidal system are quite different, as manifest in bulk diffusion of material above the film and the range and magnitude of interparticle interactions\cite{kleppmann_2017}. Nevertheless, direct measurements of island growth reveal also many similarities\cite{ganapathy_2010}, and in our previous KMC simulation the differences between a colloidal growth model (with bulk diffusion) and an SOS model were found to be small\cite{empting_2021}. Therefore, our results should also be applicable to colloidal growth, in this respect dedicated experiments with film growth of anisotropic particles are desirable.   
%Note that, while experimentally growth of colloidal systems obviously differs substantially and qualitatively from growth by evaporation in a vacuum chamber, from a simulation standpoint the difference is essentially only a quantitative one, e.g. in Energies and deposition rates\cite{ganapathy_2010,kleppmann_2017}. In this sense, our simulations are relevant for both types of experimental situations.

Summarizing, the simplified anisotropic model of this work captures some trends seen in experimental work with disc-like molecules as well as for rod-like molecules. strong roughening for rod-like molecules without an apparent strong change in molecular orientation point to the necessity of including steric interactions. Nevertheless, we think that it is useful to identify general effects of anisotropy and resulting trends on the ordering and roughness behavior of thin films using simplified, generic  models. An obvious extension of the model used in this work would be to include simple anisotropic shapes of the lattice particles, which requires more detailed considerations on the admissible moves and kinetic parameters in the system and would be computationally more expensive.   

\begin{acknowledgments}
\noindent
We thank Frank Schreiber for various insightful discussions and his helpful suggestions in finalizing this manuscript.  We gratefully acknowledge financial support of the German Research Foundation (Deutsche Forschungsgemeinschaft, DFG) through project Oe 285/3-1.
\end{acknowledgments}

\bibliography{references}

\newpage
%\clearpage

\appendix
\section{Distributions of fiber lengths}
\label{sec:appa}
\begin{figure*}[th]
	\centering
	\includegraphics[width=0.4\linewidth]{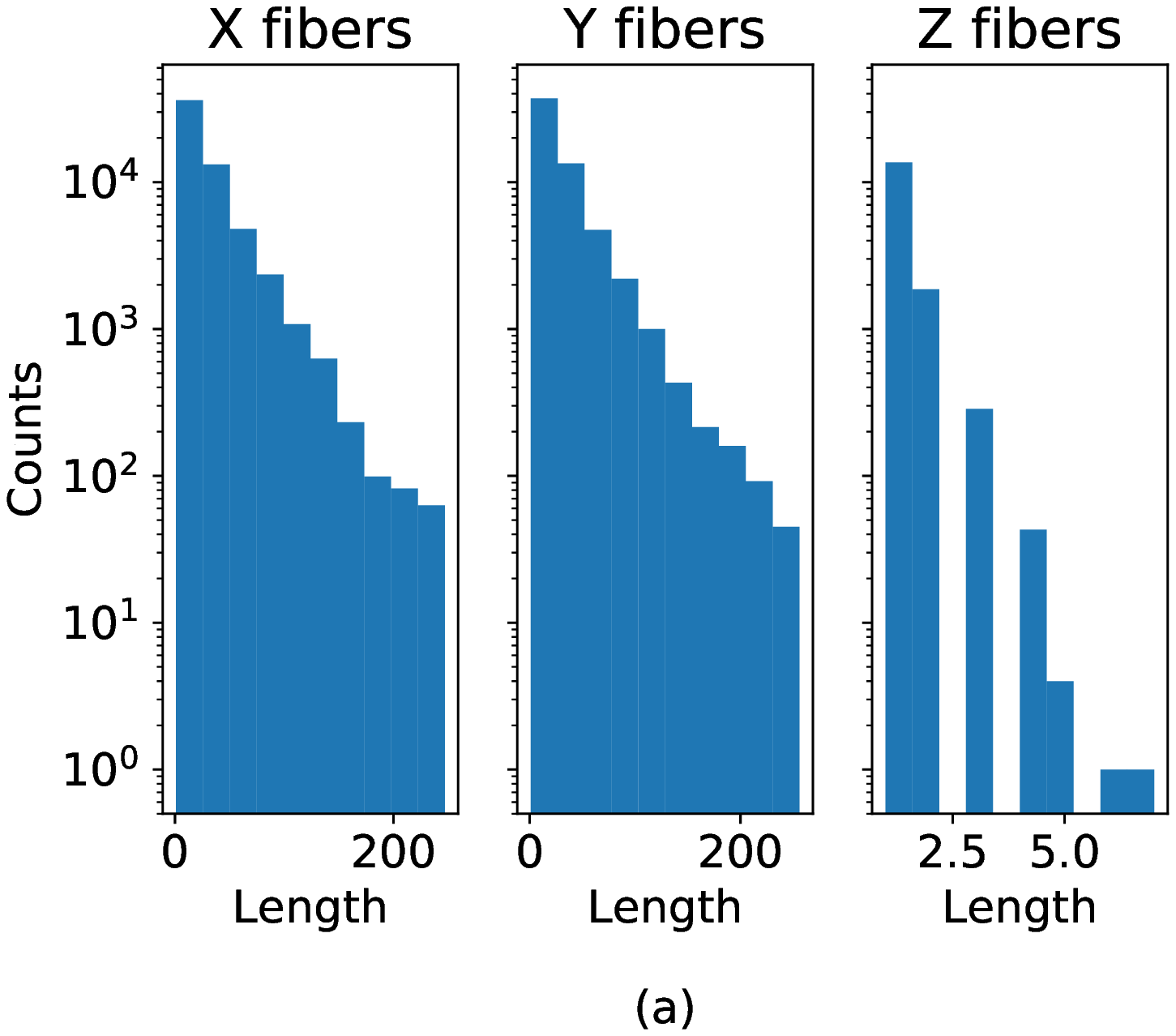}
	\includegraphics[width=0.4\linewidth]{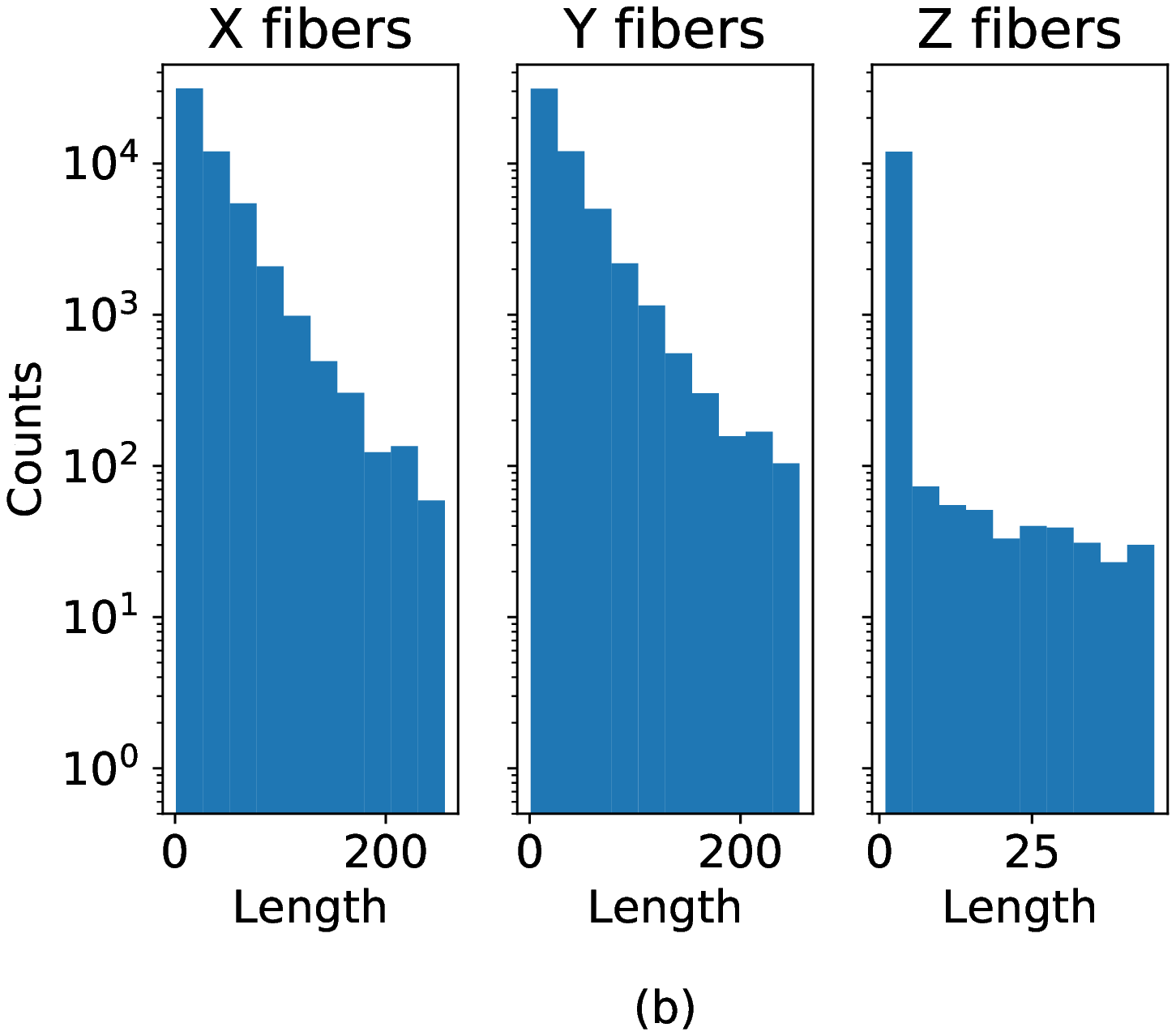}
	\caption{Fiber length distributions (a) before and (b) after the non-equilibrium transition at $\epsilon=2$, $D=10^4$, $\drot=10$ and (a) $\eta=2.6$, (b) $\eta = 2.7$}
	\label{fig:dist_jump}
\end{figure*}

In Sec. \ref{sec:noneq}, the transition points for the strong roughening transition for higher anisotropies were determined using the width of the distribution of $z$ fibers which exhibits a significant jump at the transition value $\etat$ of the non-equilibrium transition. 
In Fig. \ref{fig:dist_jump} an explicit example for the length distribution of $x$-, $y$- and $z$-fibers just before and just after the transition is shown which illustrates the origin of this behavior. 
In Fig. \ref{fig:dist_jump}(a) at $\eta < \etat$, the length distribution of $z$-fibers is narrow, not exceeding a length of $7$ while lenghths of $x$- and $y$-fibers may reach the box length and in Fig. \ref{fig:dist_jump}(b) at $\eta > \etat$, the length distribution of $z$-fibers is significantly wider, with fiber lengths going up to almost 50 (the final average film thickness). The distribution only weakly depends on the length which is also the cause for a large width of the distribution. This behavior of the $z$-fiber length distribution was found for all parameters investigated in Fig.~\ref{fig:heatmap}.

\newpage

\section{Rod-like interactions}
\label{sec:appb}
\begin{figure*}
    \centering
    \includegraphics[width=0.9\linewidth]{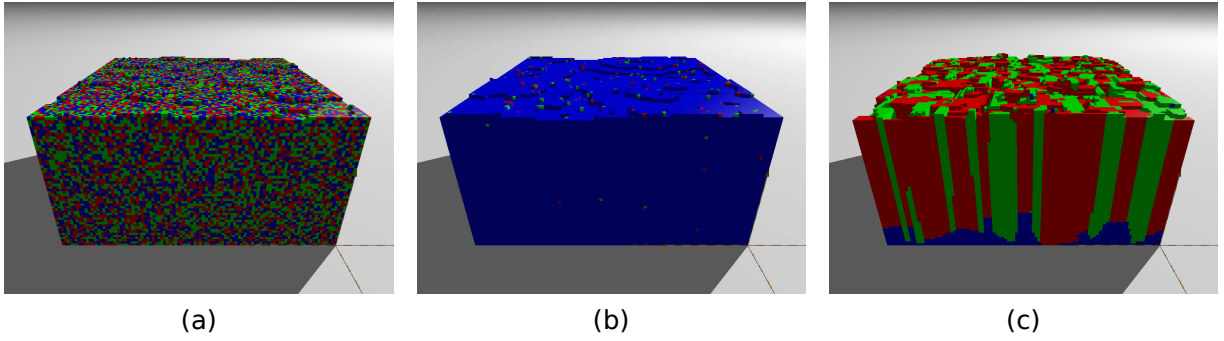}
    \caption{Snapshots of films after deposition of 50 monolayers at $\epsilon=3$, $\esub=3.11$, $D=10^4$,$\drot=100$ in model S for rod-like interactions. The colors red, green, and blue indicate that particles are oriented in the $x$, $y$ and $z$ direction, respectively. From left to right $\eta$ increases from $1$ to $1.6$ and $3$. We see how the film first goes from disordered to ordered without the morphology changing. When increasing $\eta$ to 3, $x$ and $y$ particles start forming planes, roughening the film}
    \label{fig:rod_snaps}
\end{figure*}
\begin{figure}[h]
    \centering
    \includegraphics[width=0.9\linewidth]{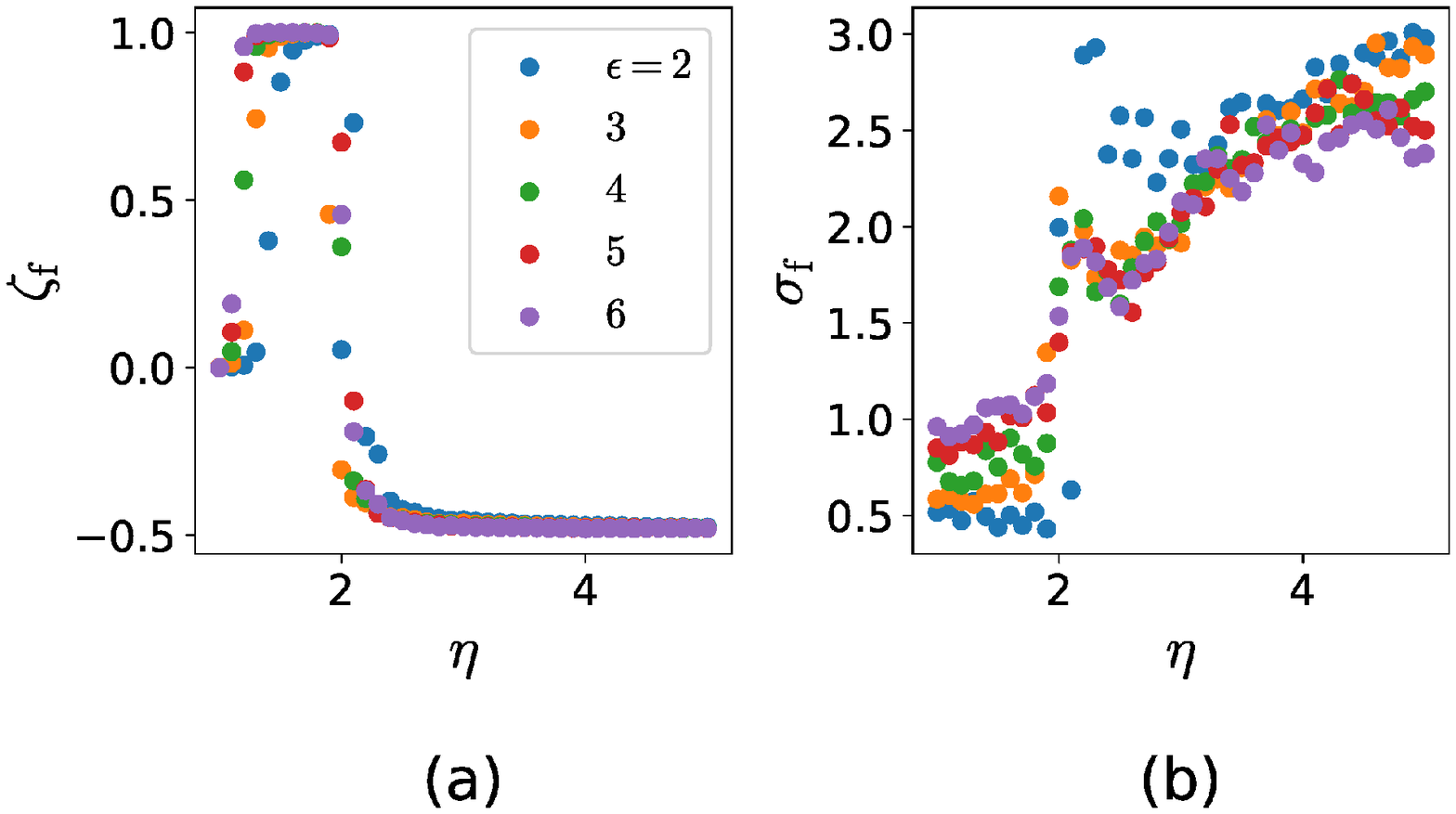}
    \caption{(a) $\zf$ and (b) $\sif$ after deposition of 50 ML vs $\eta$ at $\esub=4$, $D= 10^4$ and $\drot = 100$ for different $\epsilon$ in model S.}
    \label{fig:rods_trans}
\end{figure}

As discussed in Sec. \ref{sec:rods}, the case of interaction anisotropy emulating rod-like shapes leads to a qualitatively similar behavior as in the disc-like case. We illustrate this first with snapshots of the final film morphology for increasing anisotropy parameter, see Fig. \ref{fig:rod_snaps}. (These snapshots should be compared with
those in Fig.~\ref{fig:snapshots} for the disc-like case.) 
Fig.~\ref{fig:snapshots}(a) shows a film of isotropic particles ($\eta=1$), grown in LBL. There is a random distributions of particle orientations in the film. Fig.~\ref{fig:snapshots}(b) is for moderate anisotropy ($\eta=1.6$) the film is almost completely ordered with standing-up rods (first, equilibrium transition) but LBL growth has not changed. Fig.~\ref{fig:snapshots}(c) is for strong anisotropy ($\eta=3.0$), and $x$-,$y$-oriented particles (lying-down rods) have appeared which are energetically favored growing in perpendicular, wall-like segments. Thermal motion and the finite deposition speed lead to domains of  $x$-,$y$-oriented particles whose size is larger and whose lateral shape is more compact than found for the $z$-fibers in the disc-like case (see Fig.~\ref{fig:snapshots}(c) for a comparison). The domains lead to an increased roughness of the film (rapid roughening transition). 
%The first transition again shifts with increasing $\epsilon$.

% \begin{figure}
%     \centering
%     \includegraphics[width=\linewidth]{figures/rod_sigma.pdf}
%     \caption{$\sigma$ after deposition of 50 ML vs $\eta$ at $\esub=4$, $D= 10^4$,$\drot = 100$ for different $\epsilon$ in model S. For large $\eta$, there seems to be no difference in behavior.}
%     \label{fig:rod_sigma}
% \end{figure}

In Fig. \ref{fig:rods_trans} we show the order parameter $\zf$ and the roughness $\sif$ vs $\eta$ for different values of $\epsilon$. The transitions in $\zf$ are now opposite to those observed for disc-like interaction, i.e. the first transition is from an unordered phase to one in which all particles are oriented in the $z$ direction ($\zeta$ goes from $0$ to $1$), while at even higher values of $\eta$ the order parameter will decrease again in what appears to be a non-equilibrium transition. This is an inversion of the behavior seen for disc-like particles, but qualitatively the transitions are identical. For $\eta > \etat$, there seems to be no significant dependence of $\sif$ on $\epsilon$. $\sif$ reaches approximately the same values as in the disc-like system, due to $x$ and $y$ particles now forming wall segments, not fibers, with identically oriented particles, strongly roughening the surface (Snapshots shown in Fig. \ref{fig:rod_snaps}).

\end{document}